\documentstyle[twoside,fleqn,espcrc2,epsf]{article}

\newcommand{\AmS}{{\protect\the\textfont2
  A\kern-.1667em\lower.5ex\hbox{M}\kern-.125emS}}

\def\la{\langle}
\def\ra{\rangle}

\def\beq{\begin{equation}}
\def\eeq{\end{equation}}
\def\bea{\begin{eqnarray}}
\def\eea{\end{eqnarray}}

\def\ampl{{\mathcal{M}}}

\def\chsu3{${\mathrm{SU}}(3)_{\mathrm{L}}\times {\mathrm{SU}}(3)_{\mathrm{R}}$}
\def\lsu3{${\mathrm{SU}}(3)_{\mathrm{L}}$}
\def\rsu3{${\mathrm{SU}}(3)_{\mathrm{R}}$}

\title{$K\rightarrow\pi\pi$ matrix elements beyond the leading-order
chiral expansion
\thanks{SHEP 01/20.
Support from PPARC under grant PPA/G/S/1998/00529
is acknowledged.  We thank Sebastien Descotes and Jonathan Flynn for
discussions.}}

\author{Ph. Boucaud$^{a}$, V. Gim\'{e}nez$^{b}$, 
C.-J. D. Lin$^{c}$\thanks{Presenter at the conference.}, V. Lubicz$^{d}$, 
G. Martinelli$^{e}$, M. Papinutto$^{f}$,\\F. Rapuano$^{e}$
and C.T. Sachrajda$^{c}$ (The SPQ$_{\tiny{\mathrm{CD}}}$R Collaboration)\\
\vspace{0.3cm}
$^{a}$ Universit\'{e} de Paris Sud, L.P.T. (B\^{a}t. 210), 
       Centre d'Orsay, 91405 Orsay-Cedex, France\\
$^{b}$ Dep. de F\'{\i}sica Te\`{o}rica, IFIC,
       Dr. Moliner 50, E-46100, Burjassot, Val\`{e}ncia, Spain\\
$^{c}$ Dept. of Physics and Astronomy, Univ. of Southampton,
       Southampton SO17 1BJ, England\\
$^{d}$ Dip. di Fisica, Universit\'{a} di Roma Tre,
       Via della Vasca Navale 84, I-00146 Roma, Italy\\
$^{e}$ Dip. di Fisica, Universit\'{a} di Roma ``La Sapienza'',
       Piazzale A. Moro 2, I-00185 Roma, Italy\\
$^{f}$ Dip. di Fisica, Universit\'{a} di Pisa and INFN - Pisa, Via
       Buonarroti 2, I-56100 Pisa, Italy
}
       
\begin{document}

\begin{abstract}
We propose an approach for calculating $K\rightarrow\pi\pi$ decays
to the next-to-leading order in chiral expansion.  A detailed
numerical study of this approach is being performed.
\end{abstract}

\maketitle

\section{Introduction}
Controlling the effects of final state interactions (FSI)
is one of the main barriers towards a high-precision
theoretical prediction for $K\rightarrow\pi\pi$ decays.
This is particularly difficult for lattice QCD 
because of the analytic continuation
to Euclidean space.   The finite-volume techniques
developed in Refs. \cite{Lellouch:2000pv,Lin:2001ek} can exactly
take into account the FSI effects, but their numerical implementation
is very demanding.  In the foreseeable future,
the most practical approach,
which is also reliable and systematically improvable, 
to the study of non-leptonic kaon decays remains the
combination of lattice QCD and chiral perturbation theory ($\chi$PT).

%
%

Apart from a study of the CP-conserving, $\Delta I = 3/2$,
$K\rightarrow\pi\pi$ decay in Ref. \cite{Aoki:1998ev},
all the existing lattice results for matrix elements
of the form
$\langle\pi\pi|Q|K\rangle$ \cite{Martinelli:2001pm} 
are obtained by simulating 
matrix elements of the kind $\langle\pi|Q|K\rangle$,
then using lowest-order $\chi$PT \cite{Bernard:1985wf} to relate them to 
the desired matrix elements.  In this procedure, the effects of
higher-order chiral corrections due to FSI, 
which could be very large,
are completely missing\footnote{In Ref. \cite{Aoki:1998ev}, the decay
amplitude is also obtained at the precision of leading-order 
chiral expansion.}.

%
%

Here we present the status of an on-going project in which all the 
relevant matrix elements for $\epsilon^{\prime}/\epsilon$ and
the $\Delta I = 1/2$ rule are being calculated by computing matrix elements
of the kind $\langle\pi\pi|Q|K\rangle$, at some ``unphysical''
kinematics, on the lattice, 
and then using the chiral expansion to next-to-leading order (NLO) to
obtain them at the ``physical'' kinematics.  
In this talk, we focus\footnote{For details of 
other aspects of this work, please refer to Ref. \cite{Boucaud:2001tx}.} 
on the study of the chiral behaviour of
$\Delta I = 3/2$, 
$K\rightarrow\pi\pi$ decay amplitudes associated with the operators
\begin{eqnarray}
 Q_{4} &=&  
  (\bar{s}_{\alpha}d_{\alpha})_{L}(\bar{u}_{\beta}u_{\beta}
  - \bar{d}_{\beta}d_{\beta})_{L} + 
  (\bar{s}_{\alpha}u_{\alpha})_{L}(\bar{u}_{\beta}d_{\beta})_{L} ,\nonumber\\
\label{eq:OpDef}
 Q_{7} &=& \frac{3}{2}(\bar{s}_{\alpha}d_{\alpha})_{L}
      \sum_{q=u,d,s,c}e_{q}(\bar{q}_{\beta}q_{\beta})_{R} , \\
 Q_{8} &=& \frac{3}{2}(\bar{s}_{\alpha}d_{\beta})_{L}
      \sum_{q=u,d,s,c}e_{q}(\bar{q}_{\beta}q_{\alpha})_{R} ,\nonumber
\end{eqnarray}
where $\alpha$, $\beta$ are colour indices and $e_{q}$ is the electric
charge of $q$.  $(\bar{\psi}_{1}\psi_{2})_{L,R}$ means
$\bar{\psi}_{1}\gamma_{\mu}(1\mp\gamma_{5})\psi_{2}$.

\section{Choice of ``unphysical'' kinematics}
\label{sec:kin}
The matrix elements 
$\langle\pi^{+}\pi^{0}|Q_{i}|K^{+}\rangle$ ($i = 4, 7$ and 8)
are computed 
in the unphysical kinematics such that $K^{+}$ and one of the
pions are always at rest, while the other pion might carry 
non-zero spatial momentum.
We denote these matrix elements by
\begin{eqnarray} 
 \langle\pi^{+}(\vec{p}_{\pi})\pi^{0}(\vec{0})|Q_{i}|
  K^{+}\rangle_{\mathrm{unphys}}
 &,& \mbox{ }\mbox{ }\mathrm{if}\mbox{ }\pi^{0}
     \mbox{ }\mathrm{is}\mbox{ }\mathrm{at}\mbox{ }\mathrm{rest} ,
 \nonumber\\
  \langle\pi^{+}(\vec{0})\pi^{0}(\vec{p}_{\pi})|Q_{i}|
  K^{+}\rangle_{\mathrm{unphys}}
 &,& \mbox{ }\mbox{ }\mathrm{if}\mbox{ }\pi^{+}
     \mbox{ }\mathrm{is}\mbox{ }\mathrm{at}\mbox{ }\mathrm{rest} .\nonumber
\end{eqnarray}
Notice that the spatial momentum $\vec{p}_{\pi}$ in the above matrix 
elements might be zero as well.  In this case, both pions are at rest.
The correlators used to extract these matrix elements are discussed
in Ref. \cite{Boucaud:2001tx}.

A technical difficulty arising at this stage is that the final states
$|\pi^{+}(\vec{p}_{\pi})\pi^{0}(\vec{0})\rangle$ and 
$|\pi^{+}(\vec{0})\pi^{0}(\vec{p}_{\pi})\rangle$ are not purely
$I=2$, because Bose-Einstein statistics does not rule out the $I=1$
components.  In order to eliminate these components, we take the
symmetric combination
\begin{eqnarray}
 & &\langle\pi^{+}\pi^{0}|Q_{i}|K^{+}\rangle_{\mathrm{unphys}}^{\mathrm{sym}}
 \nonumber\\
&=& \frac{1}{2}\bigg (
\langle\pi^{+}(\vec{p}_{\pi})\pi^{0}(\vec{0})|Q_{i}|
K^{+}\rangle_{\mathrm{unphys}}\nonumber\\
& & \mbox{ }+
\langle\pi^{+}(\vec{0})\pi^{0}(\vec{p}_{\pi})|Q_{i}|
  K^{+}\rangle_{\mathrm{unphys}}
\bigg ) .
\end{eqnarray}
%

%
\section{Chiral expansion at NLO}
\label{sec:ChPT}
In order to relate
$\langle\pi^{+}\pi^{0}|Q_{i}|K^{+}\rangle_{\mathrm{unphys}}^{\mathrm{sym}}$
to their counter parts at physical kinematics, denoted as
$\langle\pi^{+}\pi^{0}|Q_{i}|K^{+}\rangle_{\mathrm{phys}}$,
we resort to NLO chiral expansion.  Under the chiral
$SU(3)_{\mathrm{L}}\otimes SU(3)_{\mathrm{R}}$ transformation, $Q_{4}$ is
in the (27,1) representation and $Q_{7,8}$ are in the (8,8) representation.
Therefore the leading term in the chiral expansion for $Q_{4}$ ($Q_{7,8}$)
is of ${\mathcal{O}}(p^{2})$ (${\mathcal{O}}(p^{0})$).  Chirally expanding
$Q_{4}$ ($Q_{7,8}$) to ${\mathcal{O}}(p^{4})$ (${\mathcal{O}}(p^{2})$)
requires the knowledge of counterterm operators in the relevant chiral
representation at this order, as well as
one-loop calculations in $\chi$PT  using the leading-order
operators.

%
%

There is only one $\chi$PT representative for operators in (27,1) 
$\big ( (8,8) \big )$ representation at ${\mathcal{O}}(p^{2})$
(${\mathcal{O}}(p^{0})$).  At ${\mathcal{O}}(p^{4})$, there are
thirty four $\chi$PT representatives for (27,1) operators, labelled as 
${\mathcal{O}}^{(27)}_{i}$ ($i=1,2,\ldots, 34$) in Ref. 
\cite{Kambor:1990tz}.  For the purpose of this work, we only need
${\mathcal{O}}^{(27)}_{2}$, ${\mathcal{O}}^{(27)}_{4}$, 
${\mathcal{O}}^{(27)}_{5}$, ${\mathcal{O}}^{(27)}_{7}$, 
${\mathcal{O}}^{(27)}_{22}$ and  ${\mathcal{O}}^{(27)}_{24}$ \footnote{Notice 
that ${\mathcal{O}}^{(27)}_{22}$ and ${\mathcal{O}}^{(27)}_{24}$ are not
included in Ref. \cite{Golterman:2000fw}. ${\mathcal{O}}^{(27)}_{24}$ also
contributes to $\langle\pi|Q|K\rangle$.}.  
The seven
${\mathcal{O}}(p^{2})$ chiral representatives for (8,8) operators
can be found in Eq. (9) of Ref. \cite{Cirigliano:1999pv}.  Here we label
them as ${\mathcal{O}}^{(8,8)}_{i}$ ($i=1,2,\ldots, 7$).

%
%

The one-loop $\chi$PT calculations at 
physical kinematics, in infinite volume, are reported in 
Ref. \cite{Golterman:1997wb} for (27,1), and 
Ref. \cite{Cirigliano:1999pv} for (8,8).
In our work, the chiral logarithms at the chosen unphysical kinematics
are much more complicated than in these references,
because of the energy-momentum injection at the operator.
The resulting amplitudes depend 
upon three kinematical
variables, $\bar{M}_{K}$ , $\bar{M}_{\pi}$ (kaon and pion masses
in the simulation) 
and $\bar{E}_{\pi}$ (energy of the pion carrying
a non-zero spatial momentum in the simulation), while they only depend upon
$M_{K}$ and $M_{\pi}$ (physical kaon and pion masses) 
at the physical kinematics.  Furthermore, our numerical
simulations are performed in a finite volume in the quenched approximation.
This requires one-loop calculations in finite-volume quenched $\chi$PT
(q$\chi$PT).  So far, we have only finished the one-loop 
(unquenched) $\chi$PT calculations in infinite volume as an exercise.
These calculations have been performed in a general way, such that 
the four-momenta of $K^{+}$, $\pi^{+}$ and $\pi^{0}$ are used as 
the kinematical variables, and energy-momentum conservation is only
implemented at the ${\mathcal{O}}(p^{2})$ strong vertices.  The desired\
kinematics is imposed at the end of the calculations.

%
%

The results for the NLO chiral expansion for the amplitudes of interest
are
\begin{eqnarray}
 & &\ampl_{\mathrm{phys}}^{(27)} = 
 \frac{-6\sqrt{2}}{f_{K}f^{2}_{\pi}} \times 
  \bigg \{  \mbox{ }
  \alpha M^{2}_{K}-
      \alpha M^{2}_{\pi}
 \nonumber\\
 &+& (\beta_{4}-\beta_{5}+4\beta_{7}+ 2 \beta_{22}) 
     M^{4}_{K}\nonumber\\
 &+& (-4\beta_{2}+3\beta_{4}+\beta_{5}-2\beta_{7}\nonumber\\
  & &\mbox{ }-2\beta_{22}
      +16\beta_{24}) M^{2}_{K}M^{2}_{\pi}\nonumber\\
 &+& (4\beta_{2}-4\beta_{4}-2\beta_{7}-16\beta_{24}) 
     M^{4}_{\pi} \mbox{ } \bigg \}  \nonumber\\
       &+& \frac{\alpha}{f_{K}f^{2}_{\pi}}\times 
      (\mathrm{one}\mbox{ }
     \mathrm{loop})^{(27)}_{\mathrm{phys}} , \nonumber\\
 & &\ampl^{(8,8)}_{\mathrm{phys}} = \frac{2\sqrt{2}}{f_{K}f^{2}_{\pi}}
  \times {\bigg \{ }
  {\gamma}\nonumber\\
  &+& [-({\delta_{2}}+\delta_{3})+
  2({\delta_{4}}+\delta_{5})+
   4{\delta_{6}}]
  {M^{2}_{K}}\nonumber\\
  &+& [({\delta_{1}}+\delta_{2})+
  4({\delta_{4}}+\delta_{5})+
  2{\delta_{6}}]
  {M^{2}_{\pi}}
  {\bigg \} }\nonumber\\
\label{eq:PhysAmp}
  &+& \frac{{\gamma}}{f_{K}f^{2}_{\pi}}
      ({\mathrm{one}\mbox{ }
     \mathrm{loop}})^{(8,8)}_{\mathrm{phys}},
\end{eqnarray}
for the physical kinematics, and
\bea
 & & (\ampl^{(27)})^{\mathrm{sym}}_{\mathrm{unphys}} =  
 \frac{-6\sqrt{2}}{f_{K}f^{2}_{\pi}} \times 
  {\bigg \{ }\nonumber\\
  & &{\alpha}(\bar{E}_{\pi}\bar{M}_{\pi}+
  \frac{\bar{E}_{\pi}\bar{M}_{K}}{2} + \frac{\bar{M}_{\pi}\bar{M}_{K}}{2})\nonumber\\
  &+& 4 \beta_{2} \bar{M}^{4}_{\pi}
  + (4\beta_{4}+2\beta_{7})\bar{E}_{\pi}\bar{M}^{3}_{\pi}\nonumber\\
  &+& (\beta_{4}-\beta_{5}+\beta_{7})\bar{M}^{3}_{\pi}\bar{M}_{K}\nonumber\\
  &+& (\beta_{4}-\beta_{5}+\beta_{7}+2\beta_{22})
              \bar{E}_{\pi}\bar{M}^{2}_{\pi}\bar{M}_{K}\nonumber\\
  &+& (-4\beta_{2}+8\beta_{24})\bar{M}^{2}_{\pi}\bar{M}^{2}_{K}\nonumber\\
  &+& (-2\beta_{5}+4\beta_{7}+4\beta_{22})
           \bar{E}_{\pi}\bar{M}_{\pi}\bar{M}^{2}_{K}\nonumber\\
  &+& (\beta_{4}+2\beta_{7})
      (\bar{M}_{\pi}+\bar{E}_{\pi})\bar{M}^{3}_{K}\nonumber\\
  &+& (-16\beta_{24})\bar{E}^{2}_{\pi}\bar{M}^{2}_{\pi}
  + 2\beta_{22}\bar{E}^{2}_{\pi} \bar{M}_{\pi} \bar{M}_{K}\nonumber\\
  &+& 8\beta_{24}\bar{E}^{2}_{\pi}\bar{M}^{2}_{K} \mbox{ } 
     {\bigg \} }\nonumber\\
       &+& \frac{{\alpha}}{f_{K}f^{2}_{\pi}}\times 
      [({\mathrm{one}\mbox{ }
     \mathrm{loop}})^{(27)}]^{\mathrm{sym}}_{\mathrm{unphys}},\nonumber\\
 & & (\ampl^{(8,8)})^{\mathrm{sym}}_{\mathrm{unphys}} = 
 \frac{2\sqrt{2}}{f_{K}f^{2}_{\pi}}
 \times {\bigg \{ }
  {\gamma}\nonumber\\
  &+& [4({\delta_{4}}+\delta_{5}) + 
 2{\delta_{6}}]{\bar{M}^{2}_{\pi}}+
 [-({\delta_{1}}+\delta_{2})] 
 {\bar{E}_{\pi}\bar{M}_{\pi}}\nonumber\\ 
 &+&[\frac{1}{2}({\delta_{1}}+\delta_{2}) - 
 ({\delta_{2}}+\delta_{3})]
      {(\bar{M}_{\pi}+\bar{E}_{\pi})\bar{M}_{K}}\nonumber\\
 &+&
 [2({\delta_{4}}+\delta_{5})+
 4{\delta_{6}}] 
 {\bar{M}^{2}_{K}}
 {\bigg \} }\nonumber\\ 
\label{eq:UnphysAmp}
 &+&\frac{{\gamma}}{f_{K}f^{2}_{\pi}}
      [({\mathrm{one}\mbox{ }
     \mathrm{loop}})^{(8,8)}]^{\mathrm{sym}}_{\mathrm{unphys}} ,
\eea
for the unphysical kinematics described in Sec. \ref{sec:kin}. Here
$\alpha$ ($\gamma$) is the coupling constant accompanying the 
${\mathcal{O}}(p^{2})$ (${\mathcal{O}}(p^{0})$) operator, and $\beta_{i}$
($\delta_{i}$) are the coupling constants associated with
${\mathcal{O}}(p^{4})$ $\big({\mathcal{O}}(p^{2})\big )$ counterterm
operators
for (27,1) $\big ( (8,8) \big )$ chiral representation.  We are
able to reproduce the 
$(\mathrm{one}\mbox{ }\mathrm{loop})^{(27)}_{\mathrm{phys}}$ and
$(\mathrm{one}\mbox{ }\mathrm{loop})^{(8,8)}_{\mathrm{phys}}$ results 
in Refs. \cite{Golterman:1997wb} and \cite{Cirigliano:1999pv}.  We have
also checked that all the amplitudes in Eqs. (\ref{eq:PhysAmp}) and
(\ref{eq:UnphysAmp}) are independent of the renormalisation scale 
$\mu_{\chi}$ in $\chi$PT with the same $\mu_{\chi}$ dependence in
each NLO coupling.

%
%

By having enough data points for the amplitudes in 
Eq. (\ref{eq:UnphysAmp}) at different values of 
$\bar{E}_{\pi}$, $\bar{M}_{\pi}$
and $\bar{M}_{K}$, one can determine the coupling constants up to
NLO chiral expansion and
construct the physical amplitudes 
in Eq. (\ref{eq:PhysAmp}).  
We are currently performing a
detailed numerical study (clover action, $\beta=6.0$) 
of this strategy, as well as the one-loop
calculations in finite-volume q$\chi$PT.  Figure 1
shows an example of our numerical work.
\begin{figure}[hbt]
\vspace{-0.9cm}
\epsfxsize=7cm
\epsfysize=5cm
\epsffile{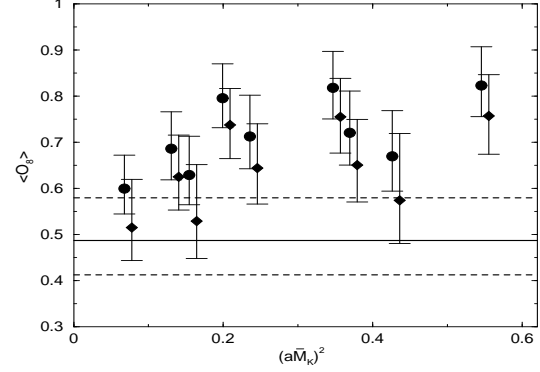}
\vspace{-1.0cm}
\label{fig:O8_plot}
\caption{{\sl Bare lattice $\la\pi\pi(I=2)|Q_{8}|K\ra$, in lattice units, 
plotted against $(a\bar{M}_{K})^{2}$. Circles are the data points in which
both pions are at rest.  Diamonds are the data points in which one of the
pions carries non-zero spatial momentum.  Points with the same 
$(a\bar{M}_{K})^{2}$ are shifted for clarity.  The solid line is the 
matrix element in the chiral limit, obtained via a fit to 
Eq. (\ref{eq:UnphysAmp}) using only data points where all the energies
are below 1 GeV and setting
chiral logs to zero. 
The dashed lines are the associated statistical errors.}}
\end{figure}

\vspace{-1.15cm}

\addcontentsline{toc}{chapter}{Bibliography}
\bibliographystyle{prsty}
\bibliography{refs}

\end{document}